\documentclass[aps,prl,twocolumn,showpacs,floatfix]{revtex4}
\usepackage{epsf}
\usepackage{subfigure}
\usepackage{amsmath}
\usepackage{amssymb}
\usepackage{graphicx}
\def\mathbi#1{\textbf{\em #1}}

\begin{document}

\title{\bf Entropy production and time asymmetry in nonequilibrium fluctuations}

\author{D. Andrieux and P. Gaspard}
\affiliation{Center for Nonlinear Phenomena and Complex Systems,\\
Universit\'e Libre de Bruxelles, Code Postal 231, Campus Plaine,
B-1050 Brussels, Belgium}

\author{S. Ciliberto, N. Garnier, S. Joubaud, and A. Petrosyan}
\affiliation{Laboratoire de Physique, CNRS UMR 5672, Ecole Normale Sup\'erieure de Lyon,
46 All\'ee d'Italie, 69364 Lyon C\'edex 07, France}

\begin{abstract}
The time-reversal symmetry of nonequilibrium fluctuations is experimentally investigated in
two out-of-equilibrium systems namely, a Brownian particle in a trap moving at constant speed
and an electric circuit with an imposed mean current.  The dynamical randomness of
their nonequilibrium fluctuations is characterized in terms of the standard and time-reversed
entropies per unit time of dynamical systems theory.  We present experimental results
showing that their difference equals the thermodynamic entropy production in units
of Boltzmann's constant.
\end{abstract}


\maketitle


Newton's equations ruling the motion of particles in matter are
known to be time-reversal symmetric. Yet, macroscopic processes
present irreversible behavior in which entropy is produced
according to the second law of thermodynamics. Recent works
suggest that this thermodynamic time asymmetry could be understood
in terms similar as those used for other symmetry breaking
phenomena in condensed matter physics. The breaking of
time-reversal symmetry should concern the fluctuations
in systems driven out of equilibrium.
These fluctuations may be described in terms of the probabilities
weighting the different possible trajectories of the systems.
Albeit the time-reversal symmetry of the microscopic Newtonian
dynamics says that each trajectory corresponds to a time-reversed
one, it turns out that distinct forward and backward trajectories
may have different probability weights if the system is out of
equilibrium. For example, the probability for a driven Brownian particle of
having a trajectory from a point  A to a point B is different of
having the same reverse trajectory from B to A.

This important observation can be further elaborated to establish
a connection with the entropy production. We consider the paths or
histories ${\mathbi z}=(z_0,z_1,z_2,...,z_{n-1})$ obtained by
sampling the trajectories $z(t)$ at regular time intervals $\tau$.
The probability weight of a typical path is known to decay as
\begin{equation}
P_+(z_0,z_1,z_2,...,z_{n-1}) \sim \exp(-n\tau h)
\label{prob}
\end{equation}
as the number $n$ of time intervals increases \cite{GP83,CP85,ER85,GW93}.
The decay rate $h$ is called the entropy per unit time
and it characterizes the temporal disorder, i.e., dynamical randomness,
in both deterministic dynamical systems
and stochastic processes \cite{GP83,CP85,ER85,GW93}.
We can compare  (\ref{prob})
with the probability weight of the time-reversed path
${\mathbi z}^{\rm R}=(z_{n-1},...,z_2,z_1,z_0)$  in the nonequilibrium system with
reversed driving constraints (denoted by the minus sign):
\begin{equation}
P_-(z_{n-1},...,z_2,z_1,z_0) \sim \exp(-n\tau h^{\rm R})\;  .
\label{prob.R}
\end{equation}
It can be shown that, out of equilibrium, the probabilities
of the time-reversed paths decay faster than the probabilities of
the paths themselves \cite{G04}. We may
interpret this as a breaking of the time-reversal symmetry in the
invariant probability distribution describing the nonequilibrium
steady state, the fundamental underlying Newtonian dynamics still
being time-reversal symmetric. The decay rate $h^{\rm R}$ in Eq.
(\ref{prob.R}) is called the time-reversed entropy per unit time
and characterizes the dynamical randomness of the time-reversed
paths \cite{G04,G05}. In the case of Markovian stochastic processes
with {\it discrete} fluctuating variables, the difference between both quantities
$h^{\rm R}$ and $h$ gives the entropy
production of irreversible thermodynamics \cite{G04,G05,LAvW05,NVdS06}.
A closely related result has been obtained for the work dissipated
in transient time-dependent systems \cite{J06}.
However, many experimental systems have {\it continuous} fluctuating variables
and evolve in nonequilibrium {\it steady} states.
Therefore, we may wonder how to measure dynamical randomness
in such systems of experimental interest and
whether the time asymmetry of this property
can be experimentally detected and related to the
thermodynamic entropy production.

In the present Letter, we provide experimental evidence for the
aforementioned time asymmetry in two nonequilibrium systems,
namely, a driven Brownian motion and a fluctuating electric circuit.
For this purpose, the decay rates $h$ and $h^{\rm R}$
are considered as so-called $(\epsilon,\tau)$-entropies per unit time,
which characterize dynamical randomness in {\it continuous-variable}
stochastic processes \cite{GW93}. These entropies per unit time
can be obtained by applying to the present stochastic systems
a method originally proposed for the study of deterministic
dynamical systems \cite{GP83,CP85,ER85}.
Thanks to this method, the $(\epsilon,\tau)$-entropies per unit time
of the paths and the corresponding time-reversed paths
can be evaluated from two long time series measured with sufficient
temporal and spatial resolutions, in two similar runs
but one driven with an opposite nonequilibrium constraint.
The experiment thus consists in recording a pair of long time series in
each system.  The dissipated heat and thermodynamic
entropy production are thus given by the difference
between the two $(\epsilon,\tau)$-entropies per unit time.

The first system is a Brownian particle dragged by an optical tweezer,
which is composed by a large aperture microscope
objective ($\times 63$, $1.3$) and by an infrared laser beam with a wavelength
of $980$ nm and a power of $20$ mW on the focal plane.
The trapped polystyrene particle has a diameter of $2$ $\mu$m and is suspended in a $20\%$
glycerol-water solution. The particle is  trapped at $20$ $\mu$m
from the bottom plate of the cell which is
$200$ $\mu$m thick.
The detection of the particle position $x_t$ is done using a  He-Ne laser
and an interferometric technique \cite{Schurr}.
In order to apply a shear to the trapped particle, the cell is moved
with a feedback-controlled piezo
which insures a perfect  linearity of  displacement.
The motion of the dragged particle is overdamped and
can be modeled as the Langevin equation
\begin{equation}
\alpha \frac{dx_{t}}{dt} = F(x_{t}-ut) + \xi(t) \; ,
\label{Langevin}
\end{equation}

where $\alpha$ is the viscous friction coefficient,
$F=-\partial_xV$ is the force exerted by the potential
$V=kx^2/2$ of the laser trap moving at constant velocity $u$,
and $\xi(t)$ a Gaussian white noise \cite{ZCC04}.
The stiffness of the potential is
$k=9.62 \; 10^{-6} \; \rm{kg} \; \rm{s}^{-2}$.
The relaxation time is $\tau_R= \alpha/k = 3.05 \; 10^{-3} \; \rm{s}$.

The second system is an electric circuit
driven out of equilibrium by a current source which imposes the mean
current $I$ \cite{GC04}. The current fluctuates in the circuit because of the
intrinsic Nyquist thermal noise \cite{ZCC04}.
The electric circuit is composed of a
capacitance $C=278 \; \rm{pF}$  in parallel with a resistance
$R=9.22 \ {\rm M}\Omega$ so that the time constant of the circuit is
$\tau_R = RC= 2.56 \; 10^{-3} \; \rm{s}$. This electric circuit
and the dragged Brownian particle, although physically different,
are known to be formally equivalent by the correspondence $\alpha
\leftrightarrow R$, $k \leftrightarrow 1/C$ and $u \leftrightarrow
I$ while the particle position $x_t$ corresponds to the charge $q_t$ inside  the
resistor at time $t$ \cite{ZCC04,GC04}.
The variables $x_t$ and $q_t$ are acquired at a sampling frequency $1/\tau=8192 \; {\rm Hz}$.

In both experiments, the temperature is $T=298\; {\rm K}$.


In order to fix the ideas, we describe our method
in the case of the dragged Brownian particle.
The heat dissipated along a
random trajectory during a time interval $t$ is given by \cite{ZCC04,S97}
\begin{equation}
Q_t =  \int_0^t \frac{dx_{t'}}{dt'} F(x_{t'}-ut') \, dt' \; .
\label{heat}
\end{equation}
After a long enough time, the system reaches
a nonequilibrium steady state, in which the entropy
production is related to the mean value of the dissipated heat
according to
\begin{equation}
 \frac{d_{\rm i}S}{dt} =
\frac{1}{T} \frac{d\langle Q_t \rangle}{dt} = \frac{\alpha u^2}{T}
\; . \label{mean}
\end{equation}

Our aim is to show that one can extract the heat dissipated
along a fluctuating path by comparing
the probability of this path, with the one of the time-reversed path having
also reversed the displacement of the potential, i.e., $u \rightarrow -u$.
We first make the change to the frame comoving with the minimum of
the potential so that $z \equiv x-ut$. After initial transients,
the system will reach a steady state characterized by a stationary
probability distribution. As we are interested in the probability
of a given succession of states corresponding to a discretization
of the signal at small time intervals $\tau$, a multi-time random
variable is defined according to ${\mathbi Z} = [Z(t_0),
Z(t_0+\tau),\ldots, Z(t_0+n\tau-\tau)]$ which corresponds to the
signal during the time period $t-t_0=n\tau$. For a stationary
process their distribution do not depend on the initial time
$t_0$. From the point of view of probability theory, the process
is defined by the $n$-time joint probabilities
$P_{\sigma}({\mathbi z}; d{\mathbi z},\tau,n) = {\rm Pr} \{
{\mathbi z} < {\mathbi Z} < {\mathbi z}+d{\mathbi z};\sigma \} =
p_{\sigma}({\mathbi z}) d{\mathbi z}$, where $p_{\sigma}({\mathbi
z})$ is the probability density for ${\mathbi Z}$ to take the
value ${\mathbi z}=(z_0,z_1,\ldots,z_{n-1})$ at times $t_0+i\tau$
for a nonequilibrium driving $\sigma=u/\vert u\vert = \pm 1$. Since
the process is Markovian, the joint probabilities
can be decomposed into the products of the Green functions
$G(z_{i},z_{i-1};\tau) dz_i$ for $i=1,\ldots,n$. $G(z,z_{0};t)$
gives the probability density for the position to be $z$ at time $t$
given that the initial position was $z_0$ \cite{C43,OM53}. To extract the
dissipation occurring along a single trajectory, one has to look
at the ratio of the probability of the forward path over the
probability of the reversed path {having also reversed the
displacement of the potential}. Indeed, taking the logarithm of
this ratio and the continuous limit $\tau \rightarrow 0$, $n
\rightarrow \infty$ with $n\tau = t$, we find
\begin{equation}
\ln \frac{P_+({\mathbi z};{\rm d} {\mathbi z},\tau,n)}{P_-({\mathbi
z}^{\rm R};{\rm d} {\mathbi z},\tau,n)}
= \beta u \int_0^t F(z_{t'}) \ dt' - \beta \Big[ V(z_t) - V(z_0) \Big]
\label{ratio}
\end{equation}
which is exactly the heat $Q_t$ in Eq. (\ref{heat})
expressed in the $z$ variable and multiplied by the inverse temperature $\beta = (k_{\rm B} T)^{-1}$.
We notice that, alone, the first term gives the work exerted by the trap \cite{S97,W02}.

Relations similar to Eq. (\ref{ratio}) have been obtained for the distribution of the work done
on a time-dependent system \cite{C99,CVK06}
and for Boltzmann's entropy production \cite{MN03}.
We emphasize that Eq. (\ref{ratio}) also holds for anharmonic potentials $V$
and that the reversal of $u$ is essential
to get the dissipated heat from the way
the path probabilities $P_+$ and $P_-$ differ.


\begin{figure}[h]
\centerline{\scalebox{0.32}{\includegraphics{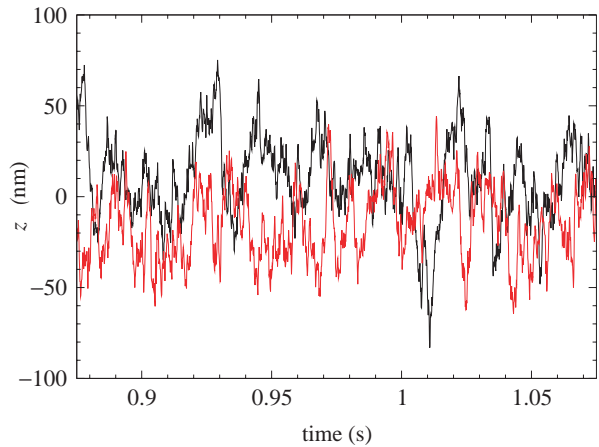}}}
\caption{Time series of typical paths $z(t)$ for the Brownian particle in the optical trap moving
at the velocity $u$ for the forward process (upper curve) and $-u$ for the reversed
process (lower curve) with $u=4.24 \ 10^{-6} m/s$.}
\label{fig1}
\end{figure}


Now, due to the continuous nature in time and in space of the
process, one has to consider $(\epsilon,\tau)$ quantities, i.e.
quantities defined on cells of size $\epsilon$ and measured at time
intervals $\tau$.
Therefore, we introduce the probability $P_+({\mathbi Z}_m;
\epsilon, \tau, n)$ for the path to remain within a distance
$\epsilon$ of some reference path ${\mathbi Z}_m$, made of $n$
successive positions of the Brownian particle observed at time
intervals $\tau$ for the forward process.  The probability is obtained
by searching for the recurrences of $M$ such reference paths or patterns in the time series.
Next, we also introduce the probability $P_-({\mathbi Z}_m^{\rm R}; \epsilon,
\tau, n)$ for a reversed path of the reversed process to remain within a distance $\epsilon$ of the
reference path ${\mathbi Z}_m$ (of the forward process) during $n$ successive
positions. According to a numerical procedure proposed by Grassberger, Procaccia and others \cite{GP83,CP85} the entropy per unit time can be estimated by the linear growth of the mean `pattern entropy' defined as
\begin{equation}
H(\epsilon,\tau,n) = - \frac{1}{M} \sum_{m=1}^M \ln
P_+({\mathbi Z}_m; \epsilon, \tau, n)
\label{pe}
\end{equation}
By similarity, we introduce
\begin{equation}
H^{\rm R} (\epsilon,\tau,n) = - \frac{1}{M} \sum_{m=1}^M \ln
P_-({\mathbi Z}_m^{\rm R}; \epsilon, \tau, n)
\label{rpe}
\end{equation}
for the reversed process.
The ($\epsilon$,$\tau$)-entropies per unit time, $h (\epsilon,\tau)$ and $h^{\rm R} (\epsilon,\tau)$, are defined by the linear growth of these mean pattern entropies as a function of the time
$n\tau$ \cite{GP83,CP85,GW93}.
In the nonequilibrium steady state, the thermodynamic entropy production
should thus be given by the difference between these two quantities:
\begin{equation}
\frac{1}{k_{\rm B}}\frac{d_{\rm i}S}{dt} = \lim_{\epsilon \rightarrow 0} \lim_{\tau
\rightarrow 0} \ [ h^{\rm R} (\epsilon,\tau) - h (\epsilon,\tau) ] \; .
\label{ds}
\end{equation}
It is important to note that the probabilities of the reversed
paths are averaged over the paths of the forward process
in order for Eq. (\ref{ds}) to hold.
The entropy production is thus expressed as the difference of two
usually very large quantities which increase with the scaling law $\epsilon^{-2}$
for $\epsilon,\tau$ going to zero \cite{GW93,G98}.
Nevertheless, their difference remains finite and gives the entropy
production in terms of the time asymmetry of the dynamical randomness
characterized by the ($\epsilon$,$\tau$)-entropies per unit time.


\begin{figure}[h]
\centerline{\scalebox{0.45}{\includegraphics{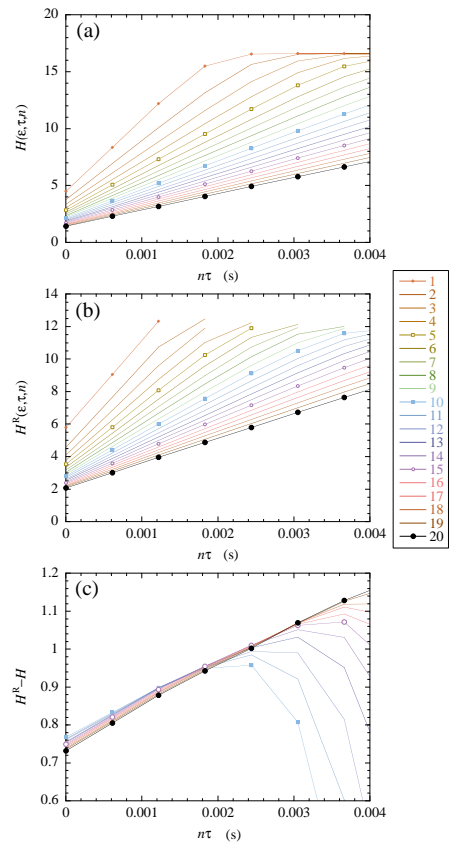}}}
\caption{(a) Entropy production of the Brownian particle versus the driving speed $u$.
The solid line is given by Eq. (\ref{mean}).
(b) Entropy production of the RC electric circuit versus the injected current $I$.
The solid line is the Joule law, $d_{\rm i}S/dt=RI^2/T$.
The dots are the results of Eq. (\ref{ds}).}
\label{fig2}
\end{figure}


In order to test experimentally that entropy production is related to
this time asymmetry according to Eq. (\ref{ds}), we have analyzed
for specific values of $\vert u \vert$ or $\vert I \vert$,
a pair of time series up to $2 \, 10^7$ points each, one  corresponding to the
forward process and the other corresponding to the reversed
process, having first discarded the transient evolution. Figure
\ref{fig1} depicts examples of paths $z(t)$ for the Brownian
particle in a moving optical trap.

For different values of $\epsilon$ between $5.6$-$11.2 \ {\rm nm}$ \cite{remark},
the mean pattern entropy (\ref{pe}) is calculated with the distance
defined by taking the maximum among the deviations
$\vert Z(t)-Z_m(t)\vert$ with respect to some reference path ${\mathbi Z}_m$
for the times $t=0,\tau,\ldots,(n-1)\tau$. The forward entropy per unit time $h(\epsilon,\tau)$
is evaluated from the linear growth of the mean
pattern entropy (\ref{pe}) with the time $n\tau$.
The backward entropy per unit time $h^{\rm R}(\epsilon,\tau)$
is obtained similarly from the time-reversed pattern entropy (\ref{rpe}).
The difference of the two dynamical entropies is depicted as in Fig. \ref{fig2}a.
The good agreement with the entropy production (\ref{mean})
is the experimental evidence that this latter is indeed related
to the time asymmetry of dynamical randomness
as predicted by Eq. (\ref{ds}).

On the other hand,  we have analyzed by the same method
the time series of the RC electric circuit.
We see in Fig. \ref{fig2}b that the entropy production
obtained from the time series analysis of the RC circuit
agrees very well with the known Joule law, which is a further confirmation
of Eq. (\ref{ds}).

\begin{figure}[h]
\centerline{\scalebox{0.42}{\includegraphics{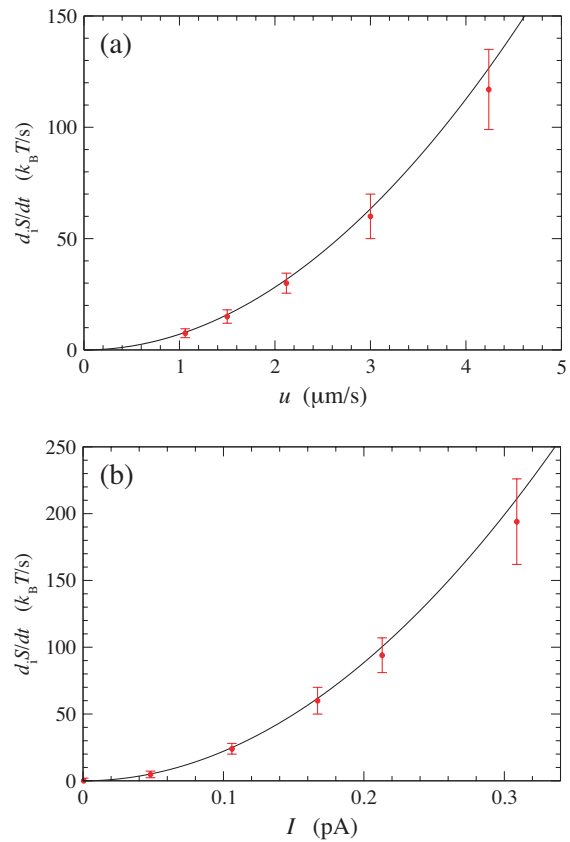}}}
\caption{Measure of the heat dissipated by the Brownian particle
along the forward and reversed paths of Fig. \ref{fig1}. The trap
velocities are $\pm u$ with $u=4.24 \ 10^{-6} m/s$. We are
searching for recurrences between the two processes. (a) Inset: A
randomly selected trajectory in the time series. The probabilities
of the corresponding forward (filled circles) and the backward (open circles) paths
for $\epsilon = 8.4\; {\rm nm}$. These probabilities present an
exponential decrease modulated by the fluctuations. (b) The
dissipated heat given by the logarithm of the ratio of the forward
and backward probabilities according to Eq. (\ref{ratio}) for
different values of $\epsilon = k \times 0.558 \ {\rm nm}$ with $k=11,\ldots,20$
in the range $6.1$-$11.2\; {\rm nm}$.
They are compared with the value (squares) directly
calculated from Eq. (\ref{heat}). For small values of $\epsilon$,
the agreement is quite good for short time
and within experimental errors for
larger time.} \label{fig3}
\end{figure}

We also tested the possibility to extract the heat (\ref{heat}) dissipated along a {\it single} stochastic path
by searching for the recurrences in the time series according to Eq. (\ref{ratio}).
A randomly selected path as well as the corresponding heat dissipated are plotted in Fig. \ref{fig3}.
We find a very good agreement so that the relation (\ref{ratio}) is also verified for single paths.
In this case, the heat exchanged between the particle and the surrounding fluid
can be positive or negative because of the molecular fluctuations.
It is only by averaging over the forward process that the dissipated heat takes
the positive value depicted in Fig. \ref{fig2}.


In conclusion, we measured the entropy production by searching
the recurrences of trajectories  in the fluctuating
dynamics of two nonequilibrium processes. The experiments we
performed consisted in the recording of two long time series. The
first one corresponds to a forward experiment while the other is
measured from the same experimental setup except that the sign of
the constraint driving the system out of equilibrium has been
reversed. From these two time series, we are able to compute two
dynamical entropies, the difference of which gives the entropy
production. Moreover, we tested the possibility to extract the
dissipated heat along a single random path. This shows that the
entropy production arises from the breaking of the time-reversal
symmetry in the probability distribution of the statistical
description of the nonequilibrium steady state. Since the decay
rates of the multi-time probabilities of the forward and reversed
paths characterize their dynamical randomness, the present results
show that the thermodynamic entropy production finds its origin
in the time asymmetry of the dynamical randomness.

{\bf Acknowledgments.} This research is financially supported by the F~.N~.R~.S~. Belgium
and the ``Communaut\'e fran\c caise de Belgique'' (contract
``Actions de Recherche Concert\'ees'' No. 04/09-312).



\begin{thebibliography}{99}
\bibitem{remark} We notice that the statistics is not sufficient for smaller values of $\epsilon$,
while the graining is too coarse for larger values.
\end{thebibliography}
\end{document}